# Deep Learning-Based Airway Segmentation in Systemic Lupus Erythematosus Patients with Interstitial Lung Disease (SLE-ILD): A Comparative High-Resolution CT Analysis

**Authors**

Sirong Piao[1*]; Ying Ming[1*]; Ruijie Zhao[1]; Jiaru Wang[1]; Ran Xiao[1]; Rui Zhao[1]; Zicheng Liao[1]; Qiqi Xu[2]; Shaoze Luo[2]; Bing Li[2]; Lin Li[2]; Zhuangfei Ma[3]; Fuling Zheng[1]; Wei Song[1#]

1 Department of Radiology, Peking Union Medical College Hospital, Chinese Academy of Medical Sciences and Peking Union Medical College, Beijing, China.

2 Research and Development Center (RDC), Canon Medical Systems (China), Beijing, China.

3 Canon Medical Systems (China), Beijing, China.

**Abstract**

Purpose: To characterize lobar and segmental airway volume differences between systemic lupus erythematosus (SLE) patients with interstitial lung disease (ILD) and those without ILD (non-ILD) using a deep learning-based approach on non-contrast chest high-resolution CT (HRCT).

Methods: A retrospective analysis was conducted on 106 SLE patients (27 SLE-ILD, 79 SLE-non-ILD) who underwent HRCT. A customized deep learning framework based on the U-Net architecture was developed to automatically segment airway structures at the lobar and segmental levels via HRCT. Volumetric measurements of lung lobes and segments derived from the segmentations were statistically compared between the two groups using two-sample t-tests (significance threshold: $p < 0.05$).

Results: At lobar level, significant airway volume enlargement in SLE-ILD patients was observed in the right upper lobe (p=0.009) and left upper lobe (p=0.039) compared to SLE-non-ILD. At the segmental level, significant differences were found in segments including R1 (p=0.016), R3 (p<0.001), and L3 (p=0.038), with the most marked changes in the upper lung zones, while lower zones showed non-significant trends.

Conclusion: Our study demonstrates that an automated deep learning-based approach can effectively quantify airway volumes on HRCT scans and reveal significant, region-specific airway dilation in patients with SLE-ILD compared to those without ILD. The pattern of involvement, predominantly affecting the upper lobes and specific segments, highlights a distinct topographic phenotype of SLE-ILD and implicates airway structural alterations as a potential biomarker for disease presence. This AI-powered quantitative imaging biomarker holds promise for enhancing the early detection and monitoring of ILD in the SLE population, ultimately contributing to more personalized patient management.

## Introduction

Systemic lupus erythematosus (SLE) is a chronic, multi-system autoimmune disease with a broad spectrum of clinical manifestations[1, 2]. Pulmonary involvement is a frequent and well-recognized complication, significantly contributing to morbidity and mortality[3]. Among these, interstitial lung disease (ILD) occurs in approximately 3–13% of individuals with SLE[4] and generally presents around a median of 6 years following the diagnosis of SLE[5]. The condition exerts a significant clinical burden: persistent hypoxia, ongoing inflammation, and hemodynamic strain accelerate disease advancement and adversely affect prognosis[6]. SLE-ILD typically manifests with a non-specific interstitial pneumonia (NSIP) pattern on histopathology and high-resolution computed tomography (HRCT), characterized by ground-glass opacities, reticulations, and traction bronchiectasis, potentially progressing to honeycombing and progressive fibrotic disease[7].

The early detection and characterization of ILD in SLE patients are critical for timely intervention and management. HRCT is the gold standard for non-invasive diagnosis and evaluation of ILD. However, conventional radiological assessment primarily focuses on parenchymal patterns of abnormality, such as the extent of fibrosis or ground-glass opacities, often relying on semi-quantitative visual scores which are subject to inter-observer variability and may lack sensitivity for detecting subtle, early changes[8, 9]. The role of the airways in SLE-ILD has received comparatively less attention. While traction bronchiectasis is a recognized feature of established fibrotic ILD, more nuanced alterations in airway dimensions, particularly in larger lobar and segmental airways, might precede overt parenchymal fibrosis or serve as indicators of specific pathophysiological processes, such as peribronchiolar inflammation or early remodeling[10].

Airway quantification, including measures of wall thickness, lumen area, and volume, has been extensively studied in obstructive airway diseases like asthma and COPD[11-13]. Its application in ILD, however, is less explored. Manual or semi-automated quantification of airway parameters is notoriously time-consuming, labor-intensive, and impractical for large-scale studies or clinical routine, especially at the segmental level. Recent advances in artificial intelligence (AI), particularly deep learning (DL), have revolutionized medical image analysis[14, 15]. Convolutional neural networks (CNNs), such as the U-Net architecture, have demonstrated exceptional performance in semantic segmentation tasks, including the precise delineation of complex tree-like structures like the bronchial tree from CT scans[16-18]. These DL-based methods offer the potential for fully automated, rapid, and highly accurate airway quantification, enabling the analysis of extensive anatomical regions that were previously infeasible to assess manually.

Currently, there is a paucity of literature applying fully automated DL-based airway segmentation to characterize SLE-ILD. The specific pattern of airway involvement at different anatomical levels (lobar, segmental) and its topographic distribution within the lungs in SLE-ILD remain poorly defined. Elucidating these patterns could provide insights into the disease pathogenesis and identify potential imaging biomarkers. In this study, we aimed to leverage a customized deep learning framework to automatically segment the airways on HRCT scans and to quantitatively characterize and compare the lobar and segmental airway volumes between SLE patients with and without ILD. We hypothesized that SLE-ILD patients would exhibit distinct regional patterns of airway volume alteration compared to their non-ILD counterparts, providing insights into the disease pathogenesis and identify potential imaging biomarkers.

## Methods and Materials

*Patients*

Our study was approved by the Institutional Review Board of our center (No. XXXX), which waived the requirement for informed consent due to the retrospective nature of the research. The participants were recruited from XXXX between 2022 and 2024, via the electronic medical records of our tertiary care institution.

The inclusion criteria were: (1) age ≥ 14 years; (2) fulfillment of the 2019 European League Against Rheumatism/American College of Rheumatology (EULAR/ACR) classification criteria for SLE[19]; and (3) availability of a non-contrast volumetric chest CT scan performed in the supine position at full inspiration. The exclusion criteria were: (1) inadequate CT image quality (significant motion artifact); (2) history of other concomitant lung diseases that could confound airway assessment (e.g., asthma, COPD, cystic fibrosis, lung cancer); (3) concurrent connective tissue disease other than SLE; (4) prior lung resection surgery; (5) active pulmonary infection at the time of CT.

*HRCT Acquisition*

The CT scans were performed on SOMATOM Force, SOMATOM Definition Flash

(Siemens Healthcare, Forchheim, Germany) and Aquilion ONE GENESIS (Cannon medical system, Tokyo, Janpan). Participants were scanned at supine position from lung apices to bases after full inspiration. The CT scan parameters and reconstruction methods were listed in **Supplementary Table 1**.

Patients enrolled were then classified into two groups: the SLE-ILD group and the SLE-non-ILD group. The HRCT images was reviewed by two experienced thoracic radiologists (with 15 and 10 years of experience, respectively). They investigated the presence of an ILD, considering ground-glass opacities, reticulations, traction bronchiectasis/bronchiolectasis and honeycombing. Our multidisciplinary team, involving pulmonologists, radiologists, pathologists and rheumatologists, confirmed all ILD diagnoses.

### *Deep Learning-Based Airway Segmentation*

We trained an automatic airway segmentation model based on the 3D-UNet network architecture. The airway segmentation pipeline first processes the CT image through lung-based cropping and intensity normalization. A 3D U-Net model then segments the airway using a topology-preserving loss function and uses sliding window inference to cover full lung. Please note that the airway at lung hilum region has been excluded. Detailed methods and results are provided in the **Supplementary material.** The final reference standard for airway was established as follows: initial deep learning segmentations were reviewed and manually corrected by two radiologists(with experience of 5 and 10 years, respectively), with any discrepancies resolved by another senior radiologist to obtain a consensus segmentation.

To further assess the spatial heterogeneity of airway at pulmonary segment level, a novel weakly supervised method, termed AHSL (Anatomy-guided Hierarchical Segmentation Learning) is applied to guide pulmonary segment segmentation[20]. This approach employs a two-stage segmentation inference strategy, using bronchovascular tree priors to guide pulmonary segment segmentation. This anatomy-driven approach aligns segment partitions without requiring pixel-level segment annotations, significantly enhancing clinical applicability. The segmentation results of pulmonary segments can yield lung regions, lobar regions, and segmental regions, all excluding the hilum.

The airway volumes of 5 different lobes (right upper, middle, lower; left upper, lower) and 18 bronchopulmonary segments were calculated automatically by the algorithm.

*Statistical Analysis*

All statistical analyses were performed using Statistical Package for the Social Sciences (SPSS) (version 26; IBM, New York, USA), and data analysis was performed using the GraphPad Prism software (version 9.5.1; GraphPad Prism software, San Diego, USA). Continuous variables are presented as mean ± standard deviation (SD) for normally distributed data. The normality of distribution for airway volume data was assessed using the Shapiro-Wilk test. Demographic and clinical characteristics between the SLE-ILD and SLE-non-ILD groups were compared using the independent samples t-test for continuous variables and the Fisher's exact test for categorical variables. Inter-group comparisons for each lobe and segment were performed using the two-sample independent t-test. A two-tailed p-value of < 0.05 was considered statistically significant. To account for the potential confounding effect of lung size, we also performed an analysis co-varying for body surface area (BSA). The DuBois formula[21] was used to calculate BSA for each subject based on their height and weight. Pulmonary airway parameters were then corrected and normalized accordingly:

$$BSA(m2) = Weight(kg)^{0.425} \times Height(cm)^{0.725} \times 0.007184$$

## Results

*Patients*

A total of 106 patients with systemic lupus erythematosus (SLE) were included in this study and were categorized into two groups: those without interstitial lung disease (SLE-non-ILD, n=79) and those with ILD (SLE-ILD, n=27). The demographic and clinical characteristics of the two groups are summarized in **Table 1**.

The two groups were comparable in terms of gender distribution (p=0.480). However, patients in the SLE-ILD group were significantly older than those in the SLE-non-ILD group (49.07 ± 13.22 years vs. 38.90 ± 13.28 years; p=0.001).

Regarding multi-system involvement, the prevalence of renal, hematologic, neuropsychiatric, and mucocutaneous manifestations was not significantly different between the two groups (all p>0.05). In contrast, musculoskeletal involvement was significantly more frequent in the SLE-ILD group compared to the SLE-non-ILD group (55.6% vs. 30.4%; p=0.023).

Respiratory symptoms such as cough, dyspnea, hemoptysis, and fever were reported infrequently and their incidence did not differ significantly between the groups (all p>0.05). Laboratory findings, including positivity for anti-ds-DNA IgG, CRP, and ESR, were also similar between SLE-non-ILD and SLE-ILD patients (all p>0.05).

*Between group differences of Lobar Airway Volumes*

Quantitative analysis of lobar airway volumes revealed significant differences between the SLE-non-ILD and SLE-ILD groups, as detailed in **Table 2**. The volume of the RUL was significantly larger in the SLE-ILD group compared to the SLE-non-ILD group (4683.04 ± 1657.47 mm³ vs. 3815.58 ± 1389.97 mm³; p=0.009). Similarly, a statistically significant increase in airway volume was observed in the LUL of SLE-ILD patients (5577.67 ± 2107.39 mm³ vs. 4725.57 ± 1724.13 mm³; p=0.039). For the other lobes, although the mean airway volumes were consistently higher in the SLE-ILD group, the differences did not reach statistical significance (p>0.05).

*Between group differences of Segmental Airway Volumes*

The comparative analysis of segmental airway volumes revealed a distinct pattern of airway dilation in the SLE-ILD group, with significant differences localized to specific segments, as detailed in **Tables 3**.

In the right lung, the SLE-ILD group exhibited significantly larger volumes in multiple segments compared to the SLE-non-ILD group. These included the apical segment (R1, p=0.016), the anterior segment (R3, p<0.001), the medial and lateral segments (R5, p=0.004), the superior segment (R6, p=0.043), and the medial basal segment (R7, p=0.009). No statistically significant differences were observed in the posterior segment (R2), the lateral segment (R4), the anterior basal segment (R8), the lateral basal segment (R9), or the posterior basal segment (R10), although a trend towards larger volume was noted in R10 (p=0.065).

In the left lung, a similar pattern was observed. Significantly larger volumes were found in the anterior segment (L3, p=0.038) and the anteromedial basal segment (L7-8, p=0.041) of the SLE-ILD patients. The volumes of the apicoposterior segment (L1-2), the superior lingular segment (L4), the inferior lingular segment (L5), the superior segment (L6), the lateral basal segment (L9), and the posterior basal segment (L10) were not significantly different between the two groups.

The results of airway segmentation at the lobar and segmental levels are displayed in **Figure 1**. The associated volume changes at the segmental level are quantified and visualized in **Figure 2**.

**Discussion**

In this study, we utilized a fully automated deep learning-based pipeline to quantitatively assess airway volumes at both lobar and segmental levels in SLE patients with and without ILD. Our principal finding is that SLE-ILD is associated with significant regional airway dilation, most pronounced in the upper lobes, specifically in several segmental bronchi within these lobes, such as R1 (apical), R3 (anterior), and L3 (anterior). Our results suggest that airway remodeling is a feature of SLE-ILD and that it follows a distinct anatomical distribution, which can be precisely quantified using AI-driven methods.

Among the pulmonary manifestations of SLE, any compartment of the lungs can be involved: the pleura (pleural effusion), lung parenchyma (acute pneumonitis, infections, and ILD), respiratory muscles, and pulmonary vessels (acute pulmonary hemorrhage)[22, 23]. Although pulmonary disease is not listed in the formal diagnostic criteria for SLE, chronic lung involvement in particular exerts a significant adverse impact on prognosis. Moreover, certain treatments for SLE are associated with an elevated risk of respiratory infections[24]. Radiographic evidence of interstitial fibrosis, manifested as a reticular pattern, mainly exhibit in the lower lobes; high-resolution CT (HRCT) reveals interstitial abnormalities in 30% of patients; these are often relatively mild and non-specific, with thickening of the interlobular and intralobular septa and parenchymal bands being the most common findings[25].

The predilection for upper lobe airway enlargement observed in our cohort is intriguing. The airway dilation we observed, particularly in the upper lobes, could represent an early manifestation of peribronchiolar inflammation and subsequent fibrosis, leading to "traction" effects on the larger airways. Alternatively, it might indicate a primary airway-centered inflammatory process specific to SLE, potentially related to immune complex deposition or lymphocytic infiltration around the bronchovascular bundles [26, 27]. The quantitative evidence provided by our study strengthens the observation that SLE-ILD may have a unique spatial phenotype.

In our study, the identification of region-specific airway dilation as a feature of SLE-ILD offers a novel potential imaging biomarker. Integrating this automated airway quantification into clinical workflows could enhance risk stratification. For instance, SLE patients showing upper lobe airway enlargement on follow-up scans might warrant closer monitoring or more aggressive management, even in the absence of significant parenchymal changes. Furthermore, this metric could serve as an objective tool for monitoring disease progression or treatment response in therapeutic trials, providing a continuous variable that is more sensitive to change than categorical visual scores. However, current research on the spatial distribution patterns of airway changes in SLE or CTD-associated lung involvement remains limited. Future longitudinal studies are warranted to investigate the evolution of these alterations in relation to disease activity and to validate findings in larger cohorts. The underlying mechanisms also require further elucidation.

From a methodological perspective, our study underscores the utility of deep learning in extracting subtle, yet potentially significant, imaging biomarkers that are difficult to assess consistently through visual evaluation alone. The ability to automatically generate precise volumetric measurements for each segmental airway represents a significant advance over previous semi-automated techniques. This objective quantification minimizes observer bias and allows for the detection of subclinical changes that might precede visible parenchymal abnormalities on HRCT, potentially aiding in the early diagnosis of ILD in asymptomatic SLE patients.

Our study has several limitations. First, its retrospective design and single-center nature may introduce selection bias, and the generalizability of our findings needs validation in larger, multi-ethnic cohorts. Second, the cross-sectional design precludes establishing a causal relationship between airway dilation and the subsequent development or progression of fibrosis; longitudinal studies are needed to determine whether these airway changes predict future functional decline. Finally, while our DL method was robust, segmentation of very peripheral airways remains challenging and might have been incomplete in some cases with severe fibrosis or image noise, though this is unlikely to have affected the measurement of lobar and segmental airways which were the focus of this study. Future research combining airway volumes with radiomic features from the surrounding parenchyma within a multivariate model could further improve the predictive power for ILD progression in SLE.

In conclusion, this study demonstrates that an automated deep learning-based approach can effectively quantify airway volumes on HRCT scans and reveal significant, region-specific airway dilation in patients with SLE-ILD compared to those without ILD. The pattern of involvement, predominantly affecting the upper lobes and specific segments, highlights a distinct topographic phenotype of SLE-ILD and implicates airway structural alterations as a potential biomarker for disease presence. This AI-powered quantitative imaging biomarker holds promise for enhancing the early detection and monitoring of ILD in the SLE population, ultimately contributing to more personalized patient management.

**Table 1 Demographic and clinical information of the patients**

|  | SLE-non-ILD (n=79) | SLE-ILD(n=27) | P value |
|---|---|---|---|
| **Gender（M/F）** | 4/75 | 2/25 | 0.480 |
| **Age（y）** | 38.90±13.28 | 49.07±13.22 | **0.001*** |
| **Multi-System Involvement in SLE** | | | |
| Renal | 35 | 12 | 1.000 |
| Hematologic | 25 | 11 | 0.481 |
| Neuropsychiatric | 18 | 5 | 0.789 |
| Mucocutaneous | 34 | 14 | 0.504 |
| Musculoskeletal | 24 | 15 | **0.023*** |
| **Respiratory System Related Symptoms** | | | |
| Hemoptysis | 2 | 0 | 0.554 |
| Fever | 8 | 2 | 0.507 |
| Dyspnea | 1 | 0 | 0.745 |
| Cough | 10 | 3 | 0.568 |
| **Laboratory Findings** | | | |
| Anti-ds-DNA IgG (+) | 37 | 10 | 0.256 |
| CRP（+） | 6 | 2 | 0.670 |
| ESR（+） | 22 | 7 | 0.530 |

**Table 2 Inter-group differences of Lobar Airway Volumes**

| Volume(mm$^3$) | SLE-non-ILD (n=79) | SLE-ILD(n=27) | P value |
|---|---|---|---|
| RUL | 3815.58±1389.97 | 4683.04±1657.47 | **0.009*** |
| RML | 2574.49±1121.00 | 3047.37±1016.74 | 0.056 |
| RLL | 6502.94±3040.40 | 7678.07±3167.45 | 0.089 |
| LUL | 4725.57±1724.13 | 5577.67±2107.39 | **0.039*** |
| LLL | 6281.54±2749.52 | 7021.04±3238.72 | 0.252 |

Notes: Right Upper Lobe, RUL; Right Middle Lobe, RML; Right Lower Lobe, RLL; Left Upper Lobe, LUL; Left Lower Lobe, LLL

**Table 3 Inter-group differences of Segmental Airway Volumes**

|  | Volume(mm$^3$) | SLE-non-ILD (n=79) | SLE-ILD(n=27) | P value |
|---|---|---|---|---|
| Right Lung | R1 | 1056.11±513.73 | 1352.04±613.62 | **0.016*** |
|  | R2 | 1351.14±572.41 | 1389.96±580.73 | 0.762 |
|  | R3 | 1408.34±540.71 | 1940.93±783.54 | **< 0.001*** |
|  | R4 | 1158.10±561.90 | 1201.19±412.07 | 0.715 |
|  | R5 | 1416.37±619.75 | 1846.11±732.86 | **0.004*** |
|  | R6 | 1316.00±547.12 | 1580.59±662.12 | **0.043*** |
|  | R7 | 1098.71±563.02 | 1432.93±576.21 | **0.009*** |
|  | R8 | 1476.22±746.43 | 1461.22±977.37 | 0.934 |
|  | R9 | 1094.38±582.37 | 1308.81±742.32 | 0.191 |
|  | R10 | 1517.61±1009.22 | 1899.63±872.59 | 0.065 |
| Left Lung | L1-2 | 1420.09±566.80 | 1590.67±546.23 | 0.176 |
|  | L3 | 1485.67±678.29 | 1813.52±755.39 | **0.038*** |
|  | L4 | 1063.18±488.52 | 1241.11±635.54 | 0.134 |
|  | L5 | 756.65±325.86 | 932.41±460.78 | 0.075 |
|  | L6 | 1358.09±580.86 | 1411.41±618.55 | 0.686 |
|  | L7-8 | 1982.39±886.38 | 2446.22±1300.96 | **0.041*** |
|  | L9 | 1215.67±799.08 | 1262.48±633.84 | 0.783 |
|  | L10 | 1725.38±1001.29 | 1901.00±1165.79 | 0.453 |

Notes:

Right Upper Lobe: R1 - Apical Segment, R2 - Posterior Segment, R3 - Anterior Segment; Right Middle Lobe: R4 - Lateral Segment, R5 - Medial Segment;

Right Lower Lobe: R6 - Superior Segment, R7 - Medial Basal Segment, R8 - Anterior Basal Segment, R9 - Lateral Basal Segment, R10 - Posterior Basal Segment;

Left Upper Lobe: L1-L2 - Apicoposterior Segment; L3 - Anterior Segment; L4 - Superior

Lingular Segment; L5 - Inferior Lingular Segment;

Left Lower Lobe: L6 - Superior Segment; L7+L8 - Anteromedial Basal Segment; L9 - Lateral Basal Segment; L10 - Posterior Basal Segment

**Figure legends**

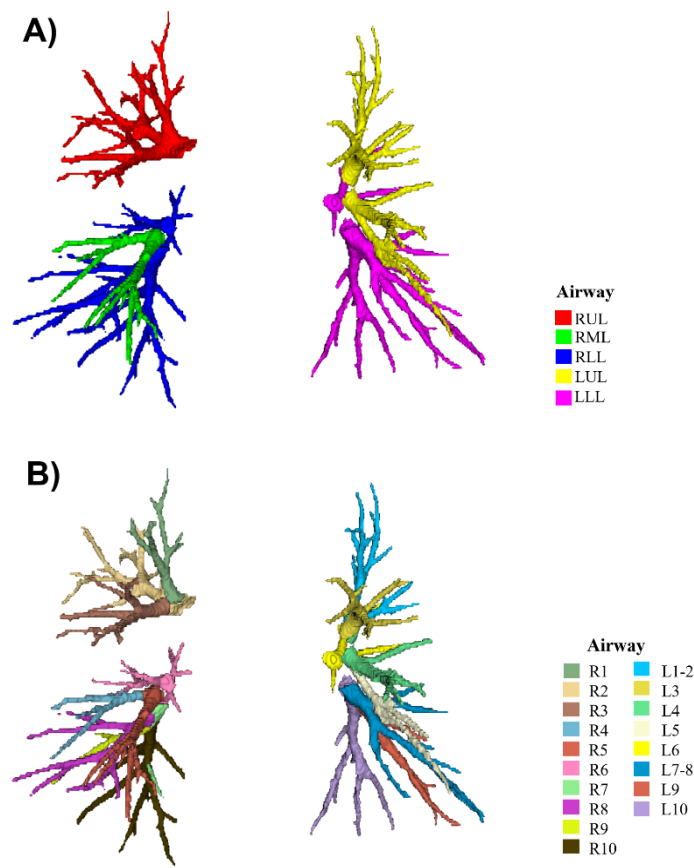

**Figure 1** Three-dimention segmentation results of pulmonary airways of different lobes(A) and segments (B).

Right Upper Lobe, RUL; Right Middle Lobe, RML; Right Lower Lobe, RLL; Left Upper Lobe, LUL; Left Lower Lobe, LLL.

Right Upper Lobe: R1 - Apical Segment, R2 - Posterior Segment, R3 - Anterior Segment; Right

Middle Lobe: R4 - Lateral Segment, R5 - Medial Segment;

Right Lower Lobe: R6 - Superior Segment, R7 - Medial Basal Segment, R8 - Anterior Basal Segment, R9 - Lateral Basal Segment, R10 - Posterior Basal Segment;

Left Upper Lobe: L1-L2 - Apicoposterior Segment; L3 - Anterior Segment; L4 - Superior Lingular Segment; L5 - Inferior Lingular Segment;

Left Lower Lobe: L6 - Superior Segment; L7+L8 - Anteromedial Basal Segment; L9 - Lateral Basal Segment; L10 - Posterior Basal Segment.

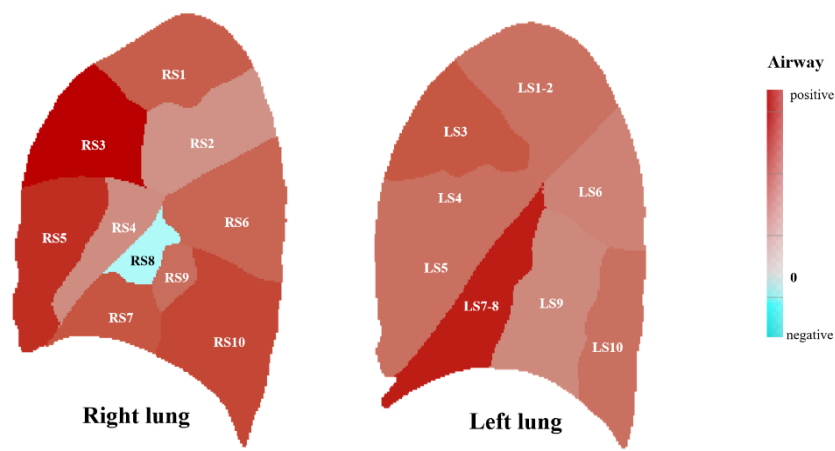

**Figure 2 Segmental level changes of airway differences in SLE-ILD population compared to SLE-non ILD group.**

Color bar indicates high and low T values (increases are shown as warm tones, decreases as cool tones)

# Supplementary material. Detailed methods and results of airway segmentation

## 1. Background and Materials information

Clinically, accurate segmentation of the pulmonary airway based on Computed Tomography (CT) is the prerequisite to the diagnosis and treatment of lung related diseases. It also plays an important role for pre-operative planning and intraoperative guidance of minimally invasive endobronchial interventions. This software provides an automatically airway segmentation algorithm through neural networks trained and tested in CT images.

In this study, we collected 256 sets of data for experiments. All experimental data were obtained from multiple institutions, including those in Japan, China, and the United States. After the data collection was completed, two experienced physicians performed the annotation of airway on the pre-segmentation results obtained from traditional thresholding algorithms. And if there is a discrepancy between the findings of two doctors, a third doctor will review and confirm the result. To ensure the model's generalizability, the experimental data included lung imaging data from NCCT and CTPA. Finally, we randomly selected 236 cases for the training set and 20 cases for the test set to train the airway segmentation model. Experimental data and GT examples are shown in Figure 1.

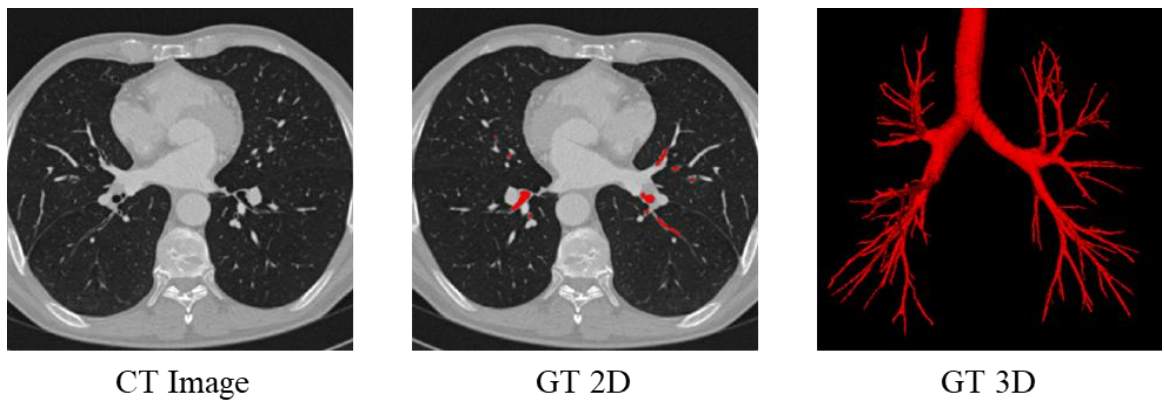

**Figure 1. Experimental CT image data and GT examples**

## 2. Segmentation model

### 2.1 Overall pipeline

The whole pipeline of airway segmentation algorithm is shown in Figure 2, including a fast lung segmentation module and an airway segmentation module. After the pulmonary CT image data is input, the model first employs a fast lung segmentation module which is trained based on a 3D U-Net architecture to rapidly obtain a coarse lung region contour. Then, the airway mask is predicted by the segmentation module, which takes the CT volume and a lung mask as inputs.

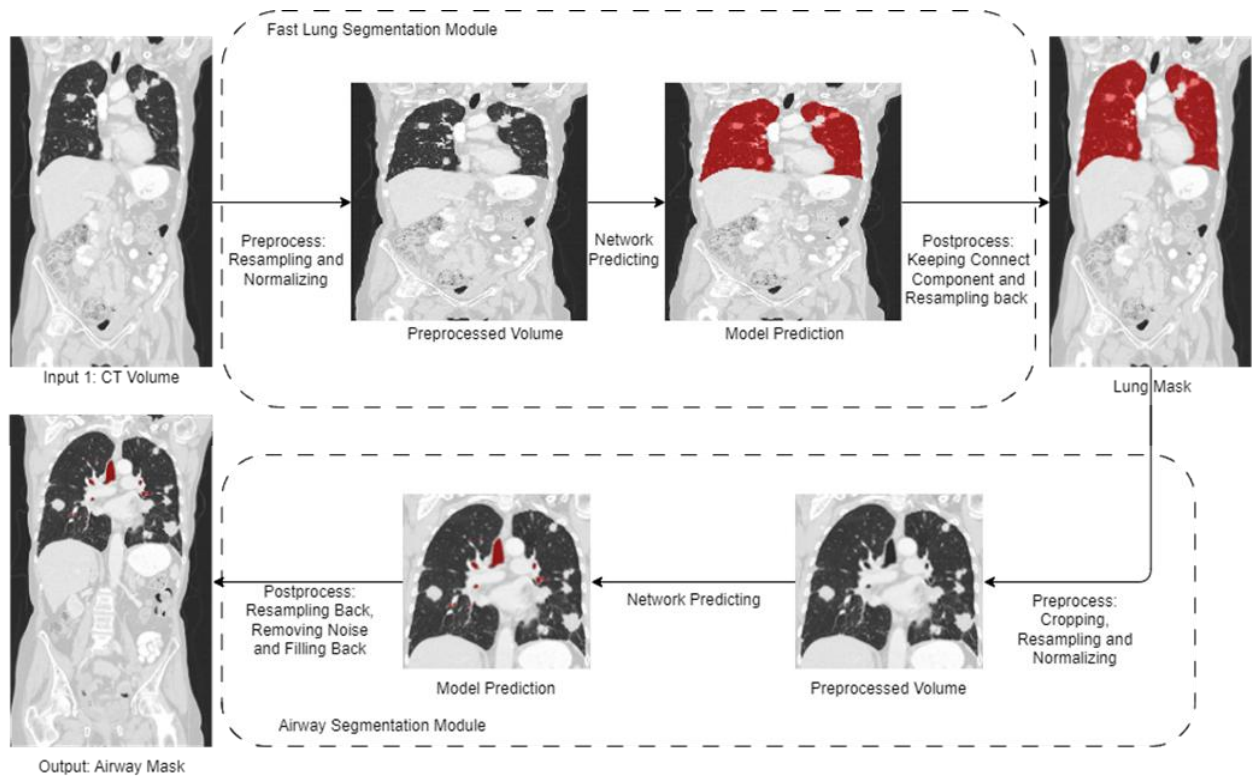

**Figure 2. Framework diagram of the airway segmentation algorithm**

### 2.2 Fast Lung Segmentation Module

In preprocessing, the input volume is resampled to fixed input size. Then its intensity is clipped into a predefined range and normalized to [0, 1]. The network predicts the input volume at once and outputs the probabilities that indicate the class (background or lung) of each voxel. In postprocessing, threshold 0.5 is used to binarize the probabilities to lung mask. The two largest connected components of lung mask are reserved, and it is resampled back to original size and spacing. The hilum is excluded from the lung mask. Finally, we trained a fast lung segmentation model—based on manually annotated lung contours and a baseline 3D U-Net architecture—to rapidly generate a coarse lung region mask, thereby defining the approximate anatomical scope for subsequent airway segmentation. The output of this component will serve as an additional input to the next segmentation network.

### 2.3 Airway Segmentation Module

We constructed the airway segmentation model using the nnUNet-v2[1]. Compared to the original baseline model, we designed the loss function as a hybrid of DICE Loss[2] and Cross-Entropy Loss (CE Loss). And the DICE loss and CE loss are weighted voxel-wise according to their distance to the airway boundary. Pixels closer to the airway boundary are assigned higher training weights to enhance the completeness and accuracy of segmentation. During training, equal weights were assigned to both loss components. Data preprocessing includes cropping, resampling, and normalization. The CT volume is cropped based on the bounding box of lung mask first. Then the network takes CT volume as input and outputs the probabilities which indicate the classes of each voxel. The classes include background and airway. From the predicted probability map, the model generates a final binary segmentation of the airway. The overall airway segmentation network architecture is shown in Figure 3.

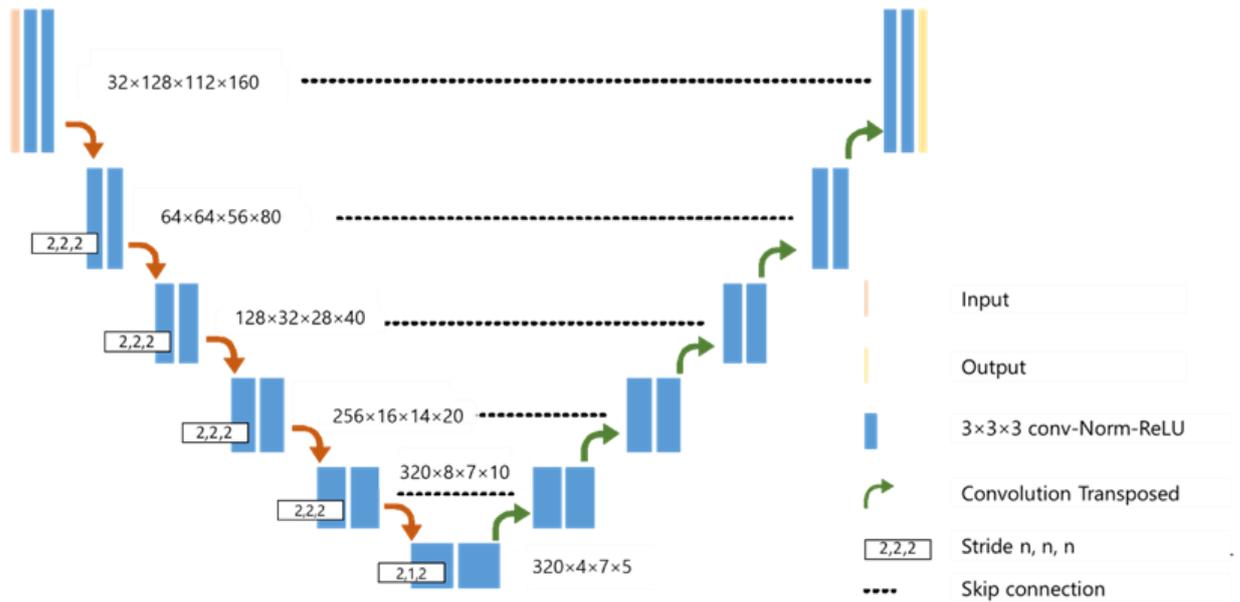

Figure 3. Schematic diagram of the Airway segmentation network structure

## 3. Evaluation method

We validated the model performance on an internal test set of 20 cases. All test cases were first meticulously annotated by experts to obtain ground truth (GT) labels. The airway segmentation model proposed in this work was then applied to automatically segment the airway in each case. The automatic segmentation results were quantitatively compared against the GT using the following evaluation metrics: DICE, Centerline Recall[3] (Cl Recall). The metric Dice is used for topological correctness measurements. Cl Recall is used to measure the topological completeness of segmentation algorithms. In addition, we also measured the inference time and memory consumption required to assess its computational efficiency.

## 4. Internal test data evaluation results

Our model was evaluated on a system equipped with an Intel(R) Xeon(R) W-2133 CPU @ 3.60 GHz, 32.0 GB of RAM (31.7 GB usable), and an NVIDIA Quadro P2000 GPU, running

Windows 10 Pro for Workstations on a 64-bit x64-based processor architecture. The average results on the 20 test cases were: DICE = 0.890 ± 0.020, Centerline Recall (Cl Recall) = 0.951 ± 0.013. These metrics indicate that the model achieves high segmentation accuracy for the airway with minimal structural discontinuities. Without using a GPU, the model achieved an average inference time of 36.8 seconds per case with memory consumption of 3.5 GB, demonstrating fast analysis speed, low memory usage, and favorable hardware efficiency.

## Supplementary Table 1. CT scan and reconstruction parameters of HRCT

|  | SOMATOM Definition Flash | SOMATOM Force | Aquilion ONE GENESIS |
|---|---|---|---|
| Scan protocol | HRCT | HRCT | HRCT |
| Tube voltage | 120kVp | 100kVp | 120kVp |
| Tube current | Automatic tube current | Automatic tube current | Automatic tube current |
| pitch | 1.2 | 1.2 | 0.813 |
| Rotation time | 0.5s | 0.5s | 0.5s |
| Image matrix | 512×512 | 512×512 | 512×512 |
| Detector configuration | 128mm×0.6mm | 192×0.6mm | 80×0.5 mm |
| Reconstruction kernel | B70f | Br40d/3 | FC52 |
| Reconstruction algorithm | SAFIRE | SAFIRE | AIDR3D standard |
| Slice thickness | 1mm | 1mm | 1mm |
| Slice interval | 1mm | 1mm | 1mm |